\documentclass{article}
\usepackage[T1]{fontenc} 
\usepackage[utf8]{inputenc} 
\usepackage{ismir,amsmath,cite,url}
\usepackage{graphicx}
\usepackage{svg}

\usepackage{color}

\usepackage[firstpage]{draftwatermark}
\definecolor{lightgray}{rgb}{0.9,0.9,0.9}
\definecolor{darkgray}{rgb}{0.4,0.4,0.4}
\SetWatermarkFontSize{12pt}
\SetWatermarkScale{1.1}
\SetWatermarkAngle{90}
\SetWatermarkHorCenter{202mm}
\SetWatermarkVerCenter{170mm}
\SetWatermarkColor{darkgray}
\SetWatermarkText{Late-Breaking / Demo Session Extended Abstract, ISMIR 2024 Conference}

\def\lbd{}

\title{Musical Source Separation of Brazilian Percussion}

\threeauthors
  {Richa Namballa} {MARL \\ New York University \\ {\tt rn2214@nyu.edu}}
  {Giovana Morais} {MARL \\ New York University \\ {\tt gv2167@nyu.edu}}
  {Magdalena Fuentes} {MARL-IDM \\ New York University \\ {\tt mf3734@nyu.edu}}

\def\authorname{R. Namballa, G. Morais, and M. Fuentes}

\begin{document}

\maketitle
\begin{abstract}
Musical source separation (MSS) has recently seen a big breakthrough in separating instruments from a mixture in the context of Western music, but research on non-Western instruments is still limited due to a lack of data. In this demo, we use an existing dataset of Brazilian \emph{samba} percussion to create artificial mixtures for training a U-Net model to separate the \emph{surdo} drum, a traditional instrument in samba. Despite limited training data, the model effectively isolates the surdo, given the drum's repetitive patterns and its characteristic low-pitched timbre. These results suggest that MSS systems can be successfully harnessed to work in more culturally-inclusive scenarios without the need of collecting extensive amounts of data.

\end{abstract}

\section{Introduction}\label{sec:introduction}

Musical source separation (MSS) is a central task of music information retrieval (MIR) which aims to ``de-mix'' audio into its corresponding instrument stems. It has applications in both the research and production of music by allowing the analysis and reuse of the stems. Most source separation systems are trained to process Western instruments only, precluding their application to more culturally-diverse music and limiting global engagement with the tools~\cite{Serra2011}.

The primary reason for the restricted usability of MSS models is the lack of diverse training data. The most commonly utilized dataset, MUSDB18, contains only four instrument categories, all of which are Western~\cite{MUSDB18}. Even those datasets which advertise a larger number of stem types are mostly focused on expanding the variety of Eurocentric instruments~\cite{Slakh, MoisesDB}. Creating new MSS datasets is challenging due to the time and monetary cost required to record and mix high-quality stems, thus the lack of diversity in instrumentation is expected. Prior to investing significant resources into constructing new datasets, we investigate the feasibility of building an MSS system by artificially creating mixtures featuring an existing non-Western dataset typically used in the context of beat tracking, the Brazilian Rhythmic Instruments Dataset (BRID)~\cite{Tomaz2016, Maia2018}.

Historically, there has been limited literature concerning the MSS task in the domain of non-Western music. Recent progress has been made in \cite{Plaja2023}, where the authors separated Carnatic vocals using training data which contained bleed in the stems by leveraging cold diffusion models. Since the audio from the BRID is very clean, we opted to utilize a simpler approach by using the standard U-Net architecture employed by both \cite{Jansson2017} and \cite{Meseguer2019} to separate the \emph{surdo}, a type of drum, from a mixture of other Brazilian percussion instruments.

\section{Data}\label{sec:data}

BRID includes two types of tracks. The first set consists of rhythmic instruments recorded as solos, with the musicians performing in various Brazilian styles using a click track. The second one contains group performances, with the musicians playing a musical style together without referencing a metronome. For this demo, we elected to set the surdo as our target source to separate from the mixture. The surdo is a large tom-like drum which plays a distinctive pattern repeated throughout the piece. This trait, plus its distinctive low-pitched timbre, makes it an easier target compared to the other percussion instruments and is thus considered a suitable starting point for this work.

\subsection{Mixture Generation}

\begin{figure*}[ht]
    \centering
    \includegraphics[width=\linewidth]{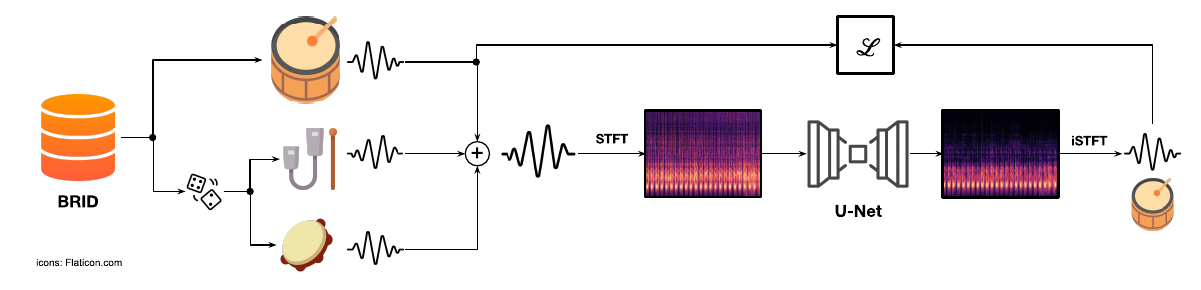}
    \caption{Overview of the training pipeline for the surdo separation model. Solo tracks are randomly selected from BRID and combined with a surdo stem to generate mixtures for training.}
    \label{fig:pipeline}
\end{figure*}

Since the group performances do not include stems, we chose to generate our own mixtures by combining the solo tracks. In total, BRID has 274 solo tracks across 10 different instruments and 5 different rhythmic styles. Of these recordings, 26 of them are surdo solos. First, we randomly allocated the surdo stems to have 22 for training, 1 for validation, and 3 for testing. Next, we split the remaining instruments into the subsets (85/5/10\%). We ensured that each instrument-style combination (e.g., \emph{tamborim-samba}) was represented in every split.

To generate a mixture within a subset, we first randomly selected a musical style. Then, we chose a surdo stem and mixed a random number of stems with it. Solos within a musical style had the same tempo and were recorded with a metronome. Therefore, the selected stems could be mixed as-is using addition, without time-alignment. The mixtures did not have any transformations applied to them (e.g., pitch-shifting or time-stretching) and there was no repetition of an instrument type within a single mixture.

In total, we generated 100 mixtures for training, 10 for validation, and 30 for testing with Scaper~\cite{Salamon2017}. We did not allow duplicate mixtures; each mixture has a unique combination of solo recordings. However, due to data limitations, there were instances of mixtures that differed only with the addition of a single stem.

\section{Methodology}\label{sec:methodology}

For this system, we used the standard 2D convolutional U-Net encoder-decoder architecture by \cite{Ronneberger2015}, adapted for operating on the frequency domain. We mimic the exact single-task implementations used in \cite{Jansson2017} and \cite{Meseguer2019} including the network architecture, hyperparameters, and spectrogram processing. We trained the surdo-separation model on the training and validation mixtures for 1000 epochs using single-GPU acceleration on an Nvidia RTX8000.

The model applies a soft mask to the input mixture's spectrogram to estimate the spectrogram of the target source. The separated audio is then reconstructed using the phase component of the original mixture and the inverse Short-Time Fourier Transform (STFT). Finally, to evaluate the performance of the model, we compute the Source-to-Distortion Ratio (SDR)~\cite{Vincent2006} for each of the test set's full mixtures, and listen to the separated audio. An overview of the pipeline is shown in Figure~\ref{fig:pipeline}.

\section{Results}

The SDR results for each data split are available in Table~\ref{tab:sdr}. The impressive SDR values indicate that the model is able to separate the surdo stem clearly. We noticed that the SDR is slightly worse on the validation set, which could be due to its small size. In future work, we plan to repeatedly sample these train / validation / test splits to account for possible biases due to the small data size.

Qualitatively, when we listen to the separated stem generated by the model, the surdo is coherent with minimal distortion. The only artifacts which occur in the stem are bleed from the other instruments in the mixture. To verify that our model is not over-fitting to the BRID solo tracks, we apply our model to the BRID group performances and even a video of a percussion ensemble from YouTube\footnote{\url{https://youtu.be/mmlK94QvwiA}}. Perceptually, our model generalizes well to these mixtures. A demo of example results is available online\footnote{\url{https://richa-namballa.github.io/mss-demo/}}.

We attribute the high SDR values and clean separation to the homogeneity of BRID and, in general, the surdo pattern rhythmic pattern. The percussion mixtures contain similar instrumentation and utilize repetitive patterns, characteristics typical of Brazilian samba. Furthermore, the surdo has a distinct timbre and frequency range which likely make it easier to isolate from the mix. Although preliminary, the fact that a simple MSS model performs a decent separation in this context is promising, as it indicates that we may not need to collect a large amount of data to achieve decent-quality separation results, as long as the music genre features certain homogeneity. While further work remains to be done to explore this hypothesis (i.e., in the remaining samba instruments), prior research in other MIR tasks indicates that this is feasible~\cite{Maia2022}.

\section{Conclusion}

In this work, we showed that we could effectively separate the surdo drum from recordings of Brazilian percussion using artificially mixed training data generated from a limited amount of stems. In future work, we hope to expand our research to experiment with other rhythmic instruments from Brazil and beyond, all in an effort to improve the cultural inclusivity of source separation applications in music.

\begin{table}
 \begin{center}
 \begin{tabular}{|c|c|c|}
  \hline
  \textbf{Dataset} & \textbf{Mean $\pm$ SD} & \textbf{Median} \\
  \hline
    Training & 16.83 $\pm$ 7.13 & 16.97 \\
  \hline
    Validation & 13.27 $\pm$ 9.61 & 12.92 \\
  \hline
    Testing & 17.57 $\pm$ 8.80 & 16.00 \\
  \hline
  \end{tabular}
\end{center}
 \caption{SDR performance of the surdo separation model.}
 \label{tab:sdr}
\end{table}

\section{Acknowledgments}
We would like to thank Google for its support in improving the representation of Latin American culture in music information research through their Award for Inclusion Research (AIR) Program.

\bibliography{ISMIR2024_lbd}

\begin{thebibliography}{10}
\providecommand{\url}[1]{#1}
\csname url@samestyle\endcsname
\providecommand{\newblock}{\relax}
\providecommand{\bibinfo}[2]{#2}
\providecommand{\BIBentrySTDinterwordspacing}{\spaceskip=0pt\relax}
\providecommand{\BIBentryALTinterwordstretchfactor}{4}
\providecommand{\BIBentryALTinterwordspacing}{\spaceskip=\fontdimen2\font plus
\BIBentryALTinterwordstretchfactor\fontdimen3\font minus \fontdimen4\font\relax}
\providecommand{\BIBforeignlanguage}[2]{{%
\expandafter\ifx\csname l@#1\endcsname\relax
\typeout{** WARNING: IEEEtran.bst: No hyphenation pattern has been}%
\typeout{** loaded for the language `#1'. Using the pattern for}%
\typeout{** the default language instead.}%
\else
\language=\csname l@#1\endcsname
\fi
#2}}
\providecommand{\BIBdecl}{\relax}
\BIBdecl

\bibitem{Serra2011}
X.~Serra, ``A multicultural approach in music information research,'' in \emph{Proceedings of the 12th International Society for Music Information Retrieval Conference}, Miami, United States, 2011, pp. 151--156.

\bibitem{MUSDB18}
\BIBentryALTinterwordspacing
Z.~Rafii, A.~Liutkus, F.-R. St{\"o}ter, S.~I. Mimilakis, and R.~Bittner, ``The {MUSDB18} corpus for music separation,'' Dec. 2017. [Online]. Available: \url{https://doi.org/10.5281/zenodo.1117372}
\BIBentrySTDinterwordspacing

\bibitem{Slakh}
E.~Manilow, G.~Wichern, P.~Seetharaman, and J.~Le~Roux, ``Cutting music source separation some {Slakh}: A dataset to study the impact of training data quality and quantity,'' in \emph{2019 IEEE Workshop on Applications of Signal Processing to Audio and Acoustics (WASPAA)}.\hskip 1em plus 0.5em minus 0.4em\relax IEEE, 2019.

\bibitem{MoisesDB}
I.~Pereira, F.~Ara{\'u}jo, F.~Korzeniowski, and R.~Vogl, ``Moises{DB}: A dataset for source separation beyond 4-stems,'' in \emph{Proceedings of the 24th International Society for Music Information Retrieval Conference}, Milan, Italy, 2023, pp. 619--626.

\bibitem{Tomaz2016}
P.~D. Tomaz~Jr., W.~S.~d. Silva~Jr., and L.~W.~P. Biscainho, ``Separação automática de instrumentos de percussão brasileira a partir de mistura pré-gravada,'' Master's thesis, Universidade Federal do Rio de Janeiro, June 2016.

\bibitem{Maia2018}
L.~S. Maia, P.~D. Tomaz~Jr., M.~Fuentes, M.~Rocamora, L.~W.~P. Biscainho, M.~V.~M. da~Costa, and S.~Cohen, ``A novel dataset of {B}razilian rhythmic instruments and some experiments in computational rhythm analysis,'' in \emph{AES Latin American Conference 2018}.\hskip 1em plus 0.5em minus 0.4em\relax Montevideo, Uruguay: Audio Engineering Society, 2018.

\bibitem{Plaja2023}
G.~Plaja-Roglans, M.~Miron, A.~Shankar, and X.~Serra, ``Carnatic singing voice separation using cold diffusion on training data with bleeding,'' in \emph{Proceedings of the 24th International Society for Music Information Retrieval Conference}, Milan, Italy, 2023, pp. 553--560.

\bibitem{Jansson2017}
A.~Jansson, E.~Humphrey, N.~Montecchio, R.~Bittner, A.~Kumar, and T.~Weyde, ``Singing voice separation with deep {U}-net convolutional networks,'' in \emph{Proceedings of the 18th International Society for Music Information Retrieval Conference}, Suzhou, China, 2017, pp. 745--751.

\bibitem{Meseguer2019}
G.~Meseguer-Brocal and G.~Peeters, ``Conditioned-{U}-net: Introducing a control mechanism in the {U}-net for multiple source separations,'' in \emph{Proceedings of the 20th International Society for Music Information Retrieval Conference}, Delft, Netherlands, 2019, pp. 159--165.

\bibitem{Salamon2017}
J.~Salamon, D.~MacConnell, M.~Cartwright, P.~Li, and J.~P. Bello, ``Scaper: A library for soundscape synthesis and augmentation,'' in \emph{2017 IEEE Workshop on Applications of Signal Processing to Audio and Acoustics (WASPAA)}.\hskip 1em plus 0.5em minus 0.4em\relax IEEE, 2017, pp. 344--348.

\bibitem{Ronneberger2015}
O.~Ronneberger, P.~Fischer, and T.~Brox, ``{U}-net: Convolutional networks for biomedical image segmentation,'' in \emph{Medical Image Computing and Computer-Assisted Intervention – MICCAI 2015, 18th International Conference, Proceedings, Part III}.\hskip 1em plus 0.5em minus 0.4em\relax Munich, Germany: Springer, 2015, pp. 234--241.

\bibitem{Vincent2006}
E.~Vincent, R.~Gribonval, and C.~F{\'e}votte, ``Performance measurement in blind audio source separation,'' \emph{IEEE Transactions on Audio, Speech, and Language Processing}, vol.~14, no.~4, pp. 1462--1469, 2006.

\bibitem{Maia2022}
L.~S. Maia, M.~Rocamora, L.~W.~P. Biscainho, and M.~Fuentes, ``Adapting meter tracking models to {L}atin {A}merican music,'' in \emph{Proceedings of the 23th International Society for Music Information Retrieval Conference}, Bengaluru, India, 2022, pp. 361--368.

\end{thebibliography}

\end{document}